\documentclass[%
 preprint,
 superscriptaddress,
 amsmath,amssymb,
 aps,
 showkeys,
 titlepage,
 endfloats*.   
]{revtex4-2}

\usepackage[utf8]{inputenc}
\usepackage[english]{babel}
\usepackage{booktabs}

\usepackage{graphicx}
\usepackage{dcolumn}
\usepackage{bm}

\usepackage[colorlinks = true,
            linkcolor = blue,
            urlcolor  = blue,
            citecolor = red,
            anchorcolor = blue]{hyperref}

\begin{document}


\title{Slow Thermalization and Long-Lived Coherence across Acoustic Phonon Branches in BAs} 

\author{Zeyu Xiang}
\thanks{These authors contributed equally.}
\affiliation{Department of Mechanical Engineering, University of California, Santa Barbara, CA 93106, USA}

\author{Ying Peng}
\thanks{These authors contributed equally.}
\affiliation{Department of Physics and Texas Center for Superconductivity at the University of Houston (TcSUH), Houston, TX 77204, USA}

\author{Ange Benise Niyikiza}
\affiliation{Department of Physics and Texas Center for Superconductivity at the University of Houston (TcSUH), Houston, TX 77204, USA}

\author{Fanghao Zhang}
\affiliation{Department of Mechanical Engineering, University of California, Santa Barbara, CA 93106, USA}

\author{Haoyuan Li}
\affiliation{SLAC National Accelerator Laboratory, Menlo Park, CA 94025, USA}

\author{Takahiro Sato}
\affiliation{SLAC National Accelerator Laboratory, Menlo Park, CA 94025, USA}

\author{Thomas Linker}
\affiliation{SLAC National Accelerator Laboratory, Menlo Park, CA 94025, USA}
\affiliation{Stanford PULSE Institute, SLAC National Accelerator Laboratory, Menlo Park, CA 94025, USA}

\author{Yanwen Sun}
\affiliation{SLAC National Accelerator Laboratory, Menlo Park, CA 94025, USA}

\author{Meredith Henstridge}
\affiliation{SLAC National Accelerator Laboratory, Menlo Park, CA 94025, USA}

\author{Vincent Esposito}
\affiliation{SLAC National Accelerator Laboratory, Menlo Park, CA 94025, USA}

\author{James F. Cotter}
\affiliation{SLAC National Accelerator Laboratory, Menlo Park, CA 94025, USA}

\author{Diling Zhu}
\affiliation{SLAC National Accelerator Laboratory, Menlo Park, CA 94025, USA}

\author{Zhifeng Ren}
\email{zren@uh.edu} \affiliation{Department of Physics and Texas Center for Superconductivity at the University of Houston (TcSUH), Houston, TX 77204, USA}

\author{Bolin Liao}
\email{bliao@ucsb.edu} \affiliation{Department of Mechanical Engineering, University of California, Santa Barbara, CA 93106, USA}

\date{\today}

\begin{abstract}
The exceptionally high thermal conductivity of cubic boron arsenide (BAs) arises from suppressed three-phonon scattering associated with its large acoustic--optical gap.
In this regime, four-phonon processes become non-negligible, creating an unusual scattering hierarchy with unexplored consequences for nonequilibrium phonon dynamics.
Here, time-resolved x-ray diffuse scattering reveals exceptionally slow, branch-dependent acoustic phonon thermalization and long-lived coherence.
Exploiting branch sensitivity in one-phonon diffuse scattering, we resolve transverse acoustic (TA) and longitudinal acoustic (LA) thermalization times of 29.2 and 13.9~ps, respectively; even the LA timescale is at least three times that in common semiconductors.
Phonon dispersion calculations assign coherent oscillations at 0.19 and 0.27~THz to the TA and LA modes, respectively; both persist with little decay over 50~ps, with comparable coherence found only in diamond.
These findings provide a direct dynamical manifestation of the weak phonon scattering underlying the exceptional thermal transport of BAs.
\end{abstract}

\maketitle


Boron arsenide (BAs) has attracted broad interest since theoretical work in 2013 predicted an exceptionally high thermal conductivity~\cite{lindsay2013first}.
This remarkable thermal transport property originates from the unusual phonon dispersion of BAs, notably the large acoustic–optical phonon gap arising from the substantial mass contrast between B and As~\cite{broido2013ab}.
The large gap strongly restricts three-phonon scattering between optical and acoustic modes, creating a bottleneck effect for phonon energy transfer~\cite{choudhry2023persistent,liu2026higher}.
Acoustic-branch bunching further suppresses scattering among acoustic modes~\cite{lindsay2015anomalous,ravichandran2020phonon,broido2013ab}.
With three-phonon scattering strongly restricted in these regimes, subsequent theoretical studies showed that four-phonon processes can contribute substantially to phonon relaxation~\cite{feng2017four,ravichandran2020phonon}, helping reconcile theoretical predictions with experimental measurements of thermal conductivity~\cite{li2018high,kang2018experimental,tian2018unusual}.
However, thermal conductivity captures only the cumulative effect of these processes near equilibrium and does not reveal how phonon scattering governs the relaxation of individual modes. 
The coexistence of three- and four-phonon scattering channels further obscures the microscopic pathways through which energy is redistributed among phonon modes.
Resolving these mode-dependent dynamics is therefore essential for identifying the characteristic pathways and timescales of nonequilibrium phonon relaxation in BAs.

Ultrafast excitation provides a direct means of probing how the unusual phonon scattering landscape of BAs governs energy redistribution far from equilibrium.
Theoretically, first-principles calculations predict strongly mode-selective relaxation, with energy initially concentrated in long-wavelength optical phonons and transferred slowly to the remaining phonon modes, challenging the conventional assumption of a thermalized phonon bath in electron-phonon two-temperature models~\cite{sadasivam2017theory}.
Experimentally, however, resolving these mode-dependent dynamics remains challenging because conventional optical probes average over phonon branches and momenta and can usually only access momenta near the zone center.
Transient reflectivity measurements, for example, reported an effective phonon-phonon relaxation time of approximately $9$~ps but could not resolve the underlying mode-specific pathways~\cite{tian2022ultraweak}. 
More recently, femtosecond stimulated Raman spectroscopy measured a room-temperature lifetime of approximately $20$~ps for zone-center optical phonons and, together with first-principles calculations, established an important role of four-phonon processes in nonequilibrium relaxation~\cite{liu2026higher}.
Beyond population relaxation, high-resolution Raman and infrared spectroscopy revealed exceptionally long coherence of zone-center optical phonons~\cite{lin2026exceptional}.
Despite these advances, direct mode-resolved measurements have remained largely confined to zone-center optical phonons, leaving the mode-dependent thermalization and coherence of acoustic phonons in BAs largely unexplored.

Enabled by advances in x-ray free-electron lasers, time-resolved x-ray diffuse scattering provides femtosecond temporal resolution together with momentum-resolved sensitivity to nonequilibrium phonon populations and lattice dynamics throughout reciprocal space~\cite{bostedt2016linac,trigo2013fourier,zhu2015phonon,teitelbaum2018direct,jiang2016origin}.
Here, we use this capability to directly resolve the nonequilibrium dynamics of acoustic phonons in BAs.
We observe unusually slow acoustic phonon thermalization on tens-of-picoseconds timescales, together with pronounced coherent dynamics.
Both the population buildup and coherent response vary strongly with direction in reciprocal space, revealing pronounced mode-dependent dynamics.
The directional dependence of the one-phonon structure factors enables us to distinguish the transverse acoustic (TA) and longitudinal acoustic (LA) branches.
The TA thermalization time is more than twice that of the LA branch, $29.2$~ps versus $13.9$~ps, while the corresponding coherent oscillations at $0.19$ and $0.27$~THz persist with little decay over $50$~ps.
As summarized in Table~\ref{tab:phonon_comparison}, even the more rapidly thermalizing LA branch in BAs thermalizes at least three times more slowly than representative semiconductors, while the persistent acoustic coherence in BAs is matched only by diamond.
These observations directly connect the unusually slow thermalization and long-lived coherence of individual acoustic branches to the weak phonon scattering underlying the exceptional thermal transport of BAs.

\begin{table*}[!htb]
\centering
\caption{
Comparison of acoustic phonon thermalization and coherence retention in BAs and representative semiconductors.
The thermalization timescale following photoexcitation is denoted by $\tau_{\rm th}$, and $A_{\mathrm{coh}}(50\,\mathrm{ps})/A_{\mathrm{coh}}(0)$ denotes the coherent amplitude retained after $50$~ps.
For the coherence comparison, acoustic phonons with comparable frequencies are selected.
}
\label{tab:phonon_comparison}

\small
\setlength{\tabcolsep}{6pt}
\renewcommand{\arraystretch}{1.15}

\begin{tabular}{lccccc}
\toprule
& \multicolumn{2}{c}{Thermalization}
& \multicolumn{3}{c}{Coherence Retention} \\

\cmidrule(lr){2-3}
\cmidrule(lr){4-6}

Material
& $\tau_{\mathrm{th}}$ (ps)
& Ref.
& $f_{\mathrm{coh}}$ (THz)
& $A_{\mathrm{coh}}(50\,\mathrm{ps})/A_{\mathrm{coh}}(0)$ 
& Ref. \\

\midrule

BAs
& $13.9-29.2$
& This work
& $0.19$--$0.26$
& No measurable decay
& This work \\

Diamond 
& $4.4$
& \cite{liu2000temperature}
& $0.06$
& No measurable decay
& \cite{maznev2018generation} \\

GaN
& $3.0$
& \cite{tsen1998time}
& $0.38$
& $71\%$
& \cite{liu2007anharmonic} \\

Si
& $2.0$
& \cite{harb2006carrier}
& $0.21$
& $42\%$
& \cite{cuffe2013lifetimes} \\

Ge
& $2.7$
& \cite{raciti2026unraveling}
& $0.09$
& $6\%$
& \cite{raciti2026unraveling} \\

PbTe
& $2.8$
& \cite{tanimura2020nonthermal}
& $0.51$
& $<1\%$
& \cite{thoen1998coherent} \\

\bottomrule
\end{tabular}
\end{table*}

\begin{figure}[!htb]
\includegraphics[width=0.9\textwidth]{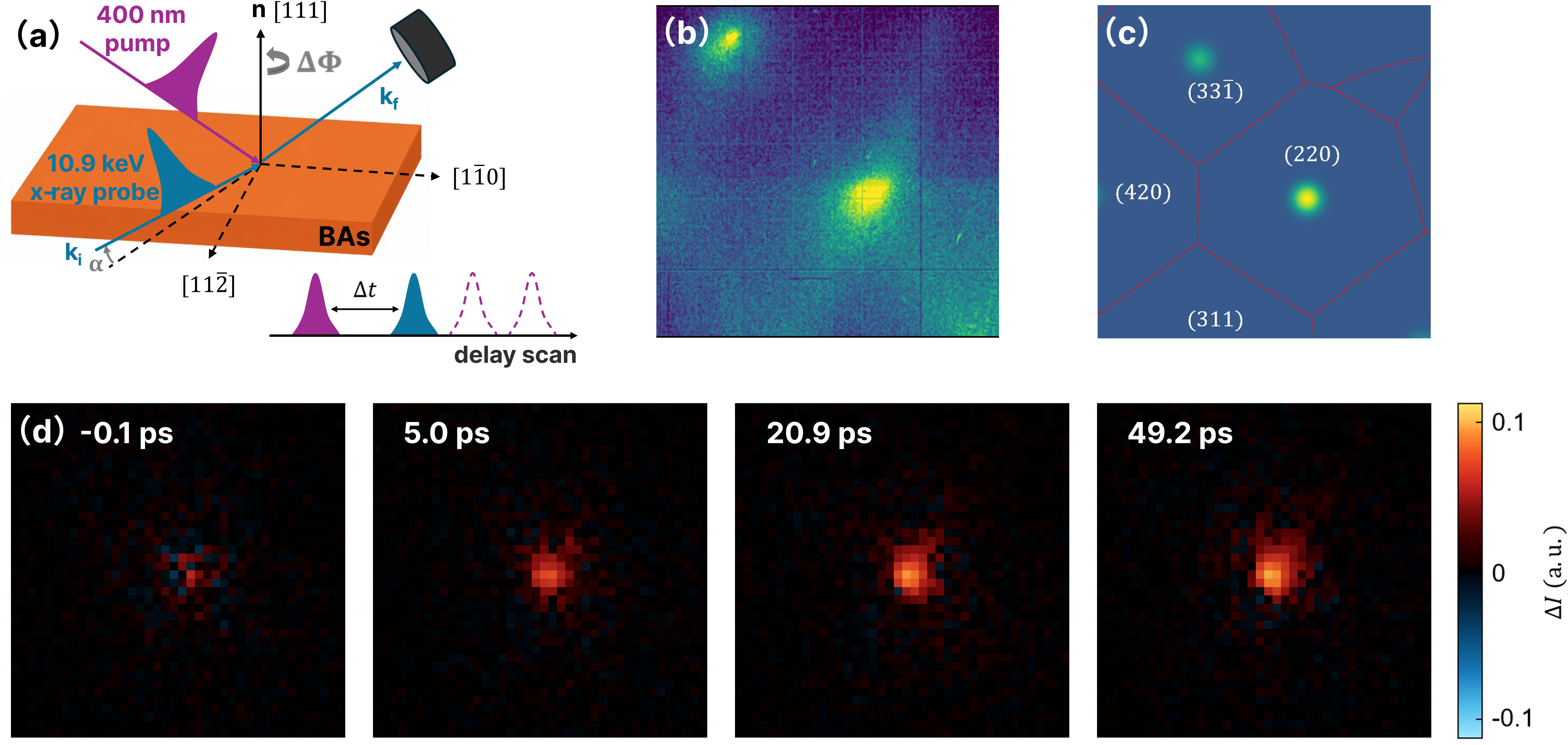}
\caption{
\textbf{Time-resolved x-ray diffuse scattering from BAs.}
(a) Optical-pump/x-ray-probe geometry using a 400-nm pump and a delayed 10.9-keV x-ray probe. The scattering vector is $\mathbf{Q}=\mathbf{k}_f-\mathbf{k}_i=\mathbf{G}_{hkl}+\mathbf{q}$, where $\mathbf{G}_{hkl}$ is a reciprocal lattice vector and $\mathbf{q}$ is the reduced wavevector.
(b) Representative pump-off diffraction pattern recorded on the detector.
(c) Calculated reciprocal-space assignment of the Bragg reflections on the detector, with the corresponding Miller indices labeled and the projected Brillouin-zone boundaries shown by dashed lines.
(d) Pump-induced differential scattering signals around the (220) reflection at selected pump--probe delays, measured at $\Delta\Phi=0.2^\circ$ away from the exact Bragg condition by rotation about the surface normal.
} 
\label{fig:fig1}
\end{figure}

We performed optical-pump/x-ray-probe scattering measurements at the x-ray pump probe (XPP) instrument of the Linac Coherent Light Source~\cite{chollet2015x} on a high-quality BAs single crystal~\cite{niyikiza2025thermal}.
Thermal conductivity mapping (Supplementary Note 1) identified a region with uniformly high thermal conductivity above 1400 W/mK at 293 K, which was selected for the XPP measurement performed at the same temperature.
Details of crystal growth and the experimental setup are provided in Methods.
As shown in Fig.~\ref{fig:fig1}(a), the BAs crystal was oriented with [111] normal to the surface and [1$\bar{1}$0] and [11$\bar{2}$] defining the surface plane; $\Delta\Phi$ denotes rotation of the sample about [111].
The sample was photoexcited by a 400 nm (3.1 eV) optical pump and probed by 10.9 keV x-ray pulses, with incident and scattered wavevectors $\mathbf{k}_{\mathrm{i}}$ and $\mathbf{k}_{\mathrm{f}}$, respectively.
The x-ray probe (optical pump) was incident on the sample at a grazing angle of $\alpha = 0.76^\circ$ ($\beta = 5.76^\circ$) to match the optical and x-ray penetration depths~\cite{henighan2016generation}.
The transient scattering dynamics were recorded using an x-ray area detector as a function of the pump--probe delay $\Delta t$.

Within the one-phonon scattering approximation~\cite{xu2005determination,warren1990x}, the diffuse scattering intensity is strongly modulated by the phonon structure factor,
\begin{equation}
I(\mathbf{Q})
\propto
\sum_j
\frac{n_j(\mathbf{q})+\frac{1}{2}}
{\omega_j(\mathbf{q})}
\left|F_j(\mathbf{Q})\right|^2 ,
\label{eq:onephonon}
\end{equation}
where $j$ denotes the phonon branch, $\mathbf{q}$ is the reduced wavevector relative to the reciprocal lattice vector $\mathbf{G}$, $n_j(\mathbf{q})$ is the phonon occupation, $\omega_j(\mathbf{q})$ is the phonon frequency, and $F_j(\mathbf{Q})$ is the one-phonon structure factor at $\mathbf{Q}=\mathbf{G}+\mathbf{q}$.
We therefore focused on the (220) Bragg reflection, where the one-phonon structure factors provide substantially greater sensitivity to acoustic than optical branches, as detailed in Supplementary Note 2.
With the optical pump blocked, the (220) reflection was aligned near the detector center at the exact Bragg condition, as shown in Fig.~\ref{fig:fig1}(b), with the corresponding reciprocal-space geometry shown in Fig.~\ref{fig:fig1}(c).
For the time-resolved measurements, the sample was detuned from the exact Bragg condition by $\Delta\Phi=0.2^\circ$ about the surface normal to suppress the Bragg contribution while retaining sensitivity to the surrounding diffuse scattering, with the corresponding rocking curve shown in Supplementary Note 3.
Under these acoustic-phonon-sensitive conditions, the pump-induced diffuse scattering near the (220) reflection exhibits an unusually slow buildup over tens of picoseconds, as illustrated by representative maps at selected pump--probe delays in Fig.~\ref{fig:fig1}(d).

\begin{figure}[!htb]
\includegraphics[width=0.9\textwidth]{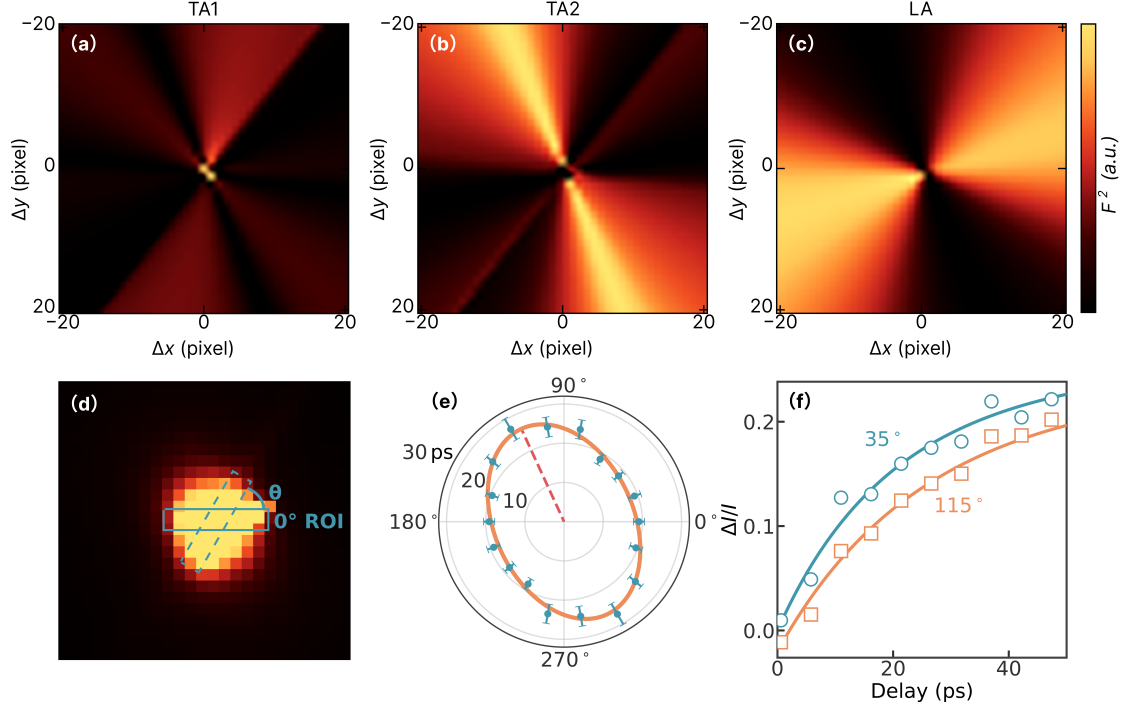}
\caption{
\textbf{Structure factors and anisotropic diffuse-scattering dynamics.}
(a)--(c) Calculated one phonon inelastic structure factors, $|F|^2$, for the TA1, TA2, and LA branches over the $40\times40$ pixel reciprocal space region sampled experimentally near the (220) Bragg reflection.
(d) Schematic of the $7\times3$ pixel ROI used for the angular analysis, rotated by an angle $\theta$ about the symmetry center; signals from symmetry related $\pm\mathbf{q}$ regions are combined in the analysis.
(e) Angular dependence of the extracted effective buildup time $\tau(\theta)$. The solid curve denotes an elliptical fit, yielding a major axis orientation of $\theta=114.7^\circ$, with $\tau_{\rm major}=26.1\pm1.8$ ps and $\tau_{\rm minor}=17.8\pm0.8$ ps.
(f) Time evolution of the diffuse scattering intensity at $\theta=35^\circ$ and $115^\circ$.
Symbols show the experimental data and solid lines the fits based on Eq.~\ref{eq:temperature_rise}; the $115^\circ$ trace is vertically offset by $-0.02$ for clarity.
The extracted thermalization times are $\tau_{\mathrm{TA}}=29.2\pm2.0$~ps and $\tau_{\mathrm{LA}}=13.9\pm1.3$~ps, demonstrating substantially slower TA thermalization.
} 
\label{fig:fig2}
\end{figure}

To determine the contributions of different acoustic branches to the measured diffuse scattering, we calculated the one-phonon structure factors $F_j(\mathbf{Q})$ for the two transverse acoustic branches (TA1 and TA2) and the longitudinal acoustic branch (LA) over the experimentally sampled region of reciprocal space, as shown in Fig.~\ref{fig:fig2}(a)--(c).
The three branches exhibit distinct directional patterns, providing branch-selective sensitivity in momentum space, while the TA1 contribution is substantially weaker than those of TA2 and LA.
The calculated patterns remain approximately inversion symmetric despite the slight detuning from the exact Bragg condition, with the symmetry center shifted slightly away from the (220) Bragg peak.
We therefore rotated a $7\times3$ pixel region of interest (ROI) about this symmetry center, as illustrated in Fig.~\ref{fig:fig2}(d), with $\theta$ varied from $0^\circ$ to $180^\circ$ in $20^\circ$ increments.
The TA2 structure factor is largest near $\theta=115^\circ$, whereas the LA structure factor is largest near $\theta=35^\circ$, providing preferential sensitivity to the two branches along distinct momentum directions.
At each orientation, signals from symmetry related $\pm\mathbf{q}$ regions were combined to improve the signal to noise ratio, retain the even one-phonon diffuse scattering component, and suppress odd contributions associated with transient lattice strain~\cite{loether2016pump,xu2005determination}.

We quantified the dynamics along these directions using the normalized diffuse scattering change $\Delta I/I=(I_{\rm on}-I_{\rm off})/I_{\rm off}$.
At each orientation, the temporal response was fitted with a single exponential rise to extract an effective buildup time $\tau(\theta)$, yielding the angular dependence shown in Fig.~\ref{fig:fig2}(e); representative fits are provided in Supplementary Note 4.
An elliptical fit yields a major axis at $\theta=114.7^\circ$, closely aligned with the TA2 structure factor maximum, while the minor axis lies near the LA maximum.
Guided by this correspondence, we selected ROIs at $\theta=35^\circ$ and $115^\circ$, near the respective LA and TA structure factor maxima, to resolve their thermalization dynamics.
Over the limited momentum range sampled by each ROI, the two TA branches are nearly degenerate and can therefore be treated together, as discussed in Supplementary Note 5.
The transient diffuse scattering response for an ROI at angle $\theta$ can then be expressed as
\begin{equation}
\frac{\Delta I}{I}(t,\theta) \propto
\sum_j
\frac{\Delta n_j(\mathbf{q},t)}
{\omega_j(\mathbf{q})}
\left|F_j(\mathbf{Q})\right|^2 .
\label{eq:temperature_rise}
\end{equation}
Fig.~\ref{fig:fig2}(f) shows the time evolution of the diffuse scattering intensity at $\theta=35^\circ$ and $115^\circ$ with a temporal bin width of $\Delta t=5$~ps, where symbols denote the experimental data and solid lines the fits based on Eq.~\ref{eq:temperature_rise}.
Assuming a Bose--Einstein occupation within each branch, we extracted the corresponding effective phonon temperature dynamics, yielding $\tau_{\mathrm{TA}}=29.2\pm2.0$~ps, more than twice the LA thermalization timescale of $\tau_{\mathrm{LA}}=13.9\pm1.3$~ps.
The measured ratio $\tau_{\mathrm{TA}}/\tau_{\mathrm{LA}}\approx2.1$ is consistent with the slower TA thermalization predicted by previous calculations~\cite{liu2026higher}, as discussed in Supplementary Note 6.
These unusually long and strongly branch-dependent thermalization times provide a direct time domain manifestation of the hot phonon bottleneck in BAs~\cite{choudhry2023persistent,liu2026higher}.

\begin{figure}[!htb]
\includegraphics[width=0.9\textwidth]{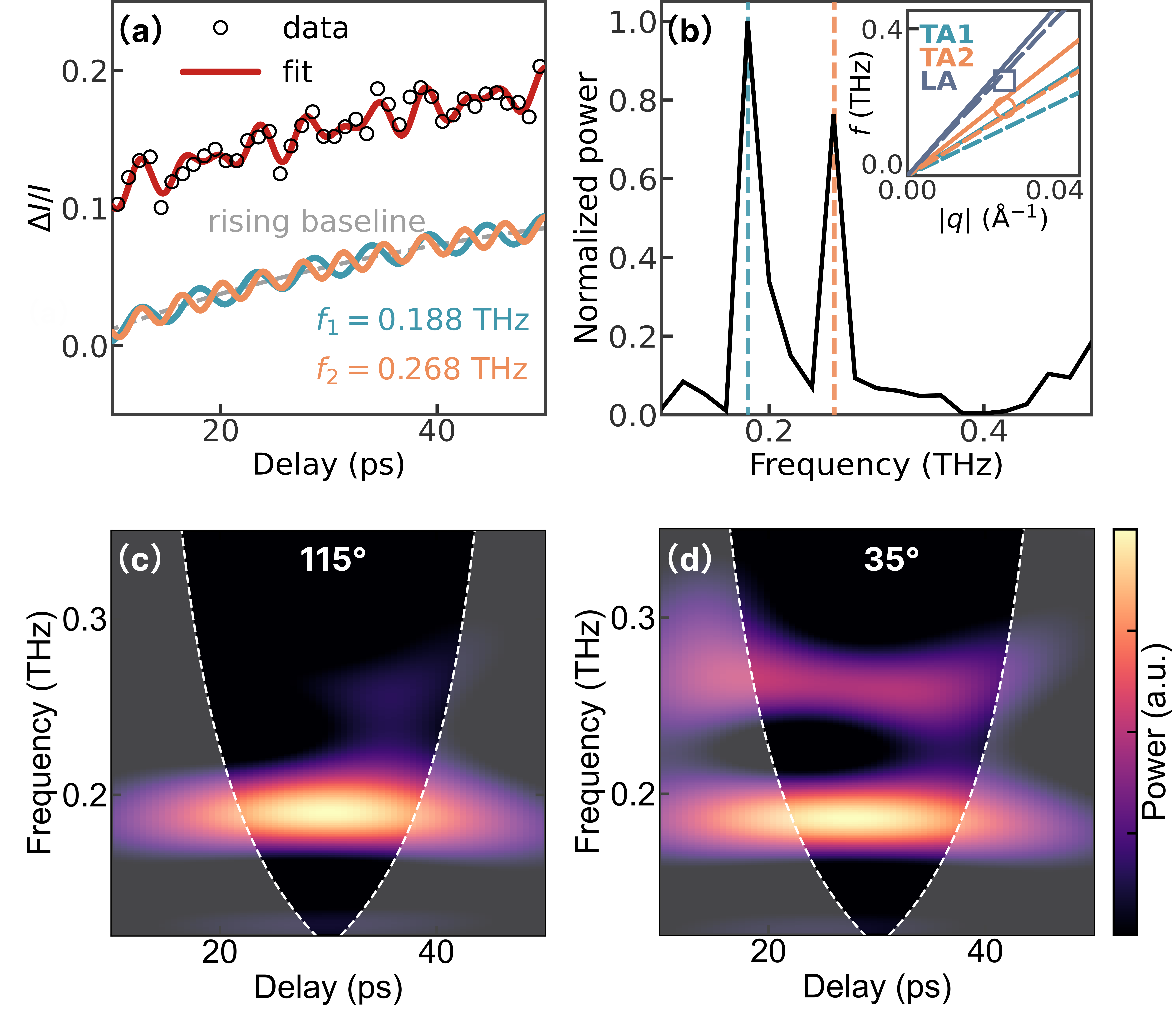}
\caption{
\textbf{Long-lived coherent acoustic-phonon dynamics.}
(a) Representative $\Delta I/I$ transient obtained with a temporal bin width of $\Delta t=1$ ps, together with a fit comprising a slowly rising background and two oscillatory components. The blue and orange curves show the extracted components at $f_1=0.188$ THz and $f_2=0.268$ THz, respectively.
(b) Normalized Fourier power spectrum of the oscillatory response, showing peaks near $0.19$ and $0.27$ THz. 
The inset compares the extracted frequencies with the TA1, TA2, and LA acoustic-phonon dispersions near the Brillouin-zone center. Solid curves denote dispersions calculated from experimentally measured elastic moduli, while dashed curves are obtained from first-principles calculations.
(c,d) Wavelet power spectra for the $\theta=115^\circ$ and $35^\circ$ ROIs, respectively. The spectral weight remains centered near $0.19$ THz throughout the measured delay range for $\theta=115^\circ$, whereas components near $0.19$ and $0.27$ THz are observed for $\theta=35^\circ$. Gray shaded regions indicate the cone of influence (COI), where edge effects become significant.
} 
\label{fig:fig3}
\end{figure}

The suppressed phonon scattering in BAs may also favor long-lived coherent acoustic motion.
As shown in Fig.~\ref{fig:fig3}(a), reducing the temporal bin width to $\Delta t=1$~ps resolves pronounced oscillations in the $\Delta I/I$ response from $10$ to $50$~ps, superimposed on the slowly rising background, whereas the first $10$~ps are dominated by the initial buildup with no well-defined periodic oscillation resolved (Supplementary Note 7).
Fourier analysis reveals two prominent components near $0.19$ and $0.27$~THz, as shown in Fig.~\ref{fig:fig3}(b).
We therefore fit the transient response with a single exponential rise and two damped oscillatory components,
\begin{equation}
\frac{\Delta I}{I}(t)=
A\left(1-e^{-t/\tau_{\rm rise}}\right)
+
\sum_{i=1}^{2}
A_i e^{-t/\tau_i}
\cos\left(2\pi f_i t+\phi_i\right),
\end{equation}
where $\tau_{\rm rise}$ denotes the population buildup time discussed above, and $\tau_i$, $f_i$, $A_i$, and $\phi_i$ denote the decay time, frequency, amplitude, and phase of the $i$th coherent component, respectively.
The fit yields $f_1=0.188$~THz and $f_2=0.268$~THz, consistent with the Fourier analysis and robust against temporal binning, as detailed in Supplementary Note 8.
At the experimentally sampled wavevectors, these frequencies agree with the TA and LA dispersions, respectively, as shown in the inset of Fig.~\ref{fig:fig3}(b), where the solid and dashed lines denote dispersions obtained from the measured elastic moduli~\cite{mahat2021elastic} and first-principles calculations, respectively.
Remarkably, neither coherent component exhibits measurable decay over the $50$~ps observation window, indicating exceptionally weak dephasing.

The directional selectivity provided by the one-phonon structure factors enables an independent test of the mode assignment and coherence persistence.
We therefore applied a Morlet wavelet transform~\cite{torrence1998practical} to the $\theta=115^\circ$ and $35^\circ$ ROI signals, which provide the strongest contrast between the TA and LA contributions, with the wavelet formalism described in Supplementary Note 9.
The $\theta=115^\circ$ spectrum in Fig.~\ref{fig:fig3}(c) is dominated by the $0.19$~THz component, consistent with the strong TA sensitivity along this direction.
By contrast, the $\theta=35^\circ$ spectrum in Fig.~\ref{fig:fig3}(d) contains both the dominant $0.19$~THz component and a weaker feature near $0.27$~THz, reflecting the enhanced LA sensitivity.
The weaker LA response despite its large structure factor follows from the inverse frequency weighting of the one-phonon diffuse scattering intensity.
Within the region unaffected by the cone of influence denoted by the shaded region in Fig.~\ref{fig:fig3}(c) and (d), the wavelet power remains nearly constant with delay, independently confirming the persistence of the coherent response.
Together, the slow branch dependent thermalization and persistent acoustic coherence provide a direct dynamical signature of the unusually restricted phonon scattering underlying the exceptional thermal conductivity of BAs.

The broader comparison in Table~\ref{tab:phonon_comparison} places these dynamics in the context of representative semiconductors.
BAs stands out in its exceptionally slow thermalization, with $\tau_{\rm th}$ ranging from $13.9$ to $29.2$~ps for the LA and TA branches, substantially longer than the few-picosecond timescales reported for diamond, GaN, Si, Ge, and PbTe.
The contrast is equally striking in the coherent response.
BAs and diamond lie at one extreme, with no measurable decay of their coherent acoustic modes over $50$~ps, whereas PbTe lies at the other, retaining less than $1\%$ of its initial coherent amplitude; GaN, Si, and Ge retain $71\%$, $42\%$, and $6\%$, respectively.
This trend broadly parallels their thermal transport properties: BAs and diamond exhibit the highest thermal conductivities among the materials considered, whereas PbTe has the lowest~\cite{yuan2023lattice}.
Although differences in phonon frequency and experimental conditions preclude a quantitative correlation, this correspondence points to a common role of weak phonon scattering in sustaining both long-lived acoustic coherence and efficient thermal transport.

Together, these results reveal the far-from-equilibrium signatures of restricted phonon scattering in BAs: unusually slow and strongly branch-dependent acoustic phonon thermalization together with persistent acoustic coherence.
They provide a direct time-domain link between the dynamics of individual acoustic phonon branches and the scattering physics underlying the exceptional thermal transport of BAs.


\section*{METHODS}
\subsection*{Crystal growth}
High-quality single-crystalline boron arsenide (BAs) was prepared through a four-step growth procedure.
First, the quartz tube was thoroughly cleaned by sequential ultrasonication in acetone, ethanol, and deionized water, followed by oven drying to remove residual moisture.
The arsenic source was further purified to reduce trace contaminants, including Si, C, and O~\cite{fedorov2021processes}.
Second, purified arsenic ($>99.9999\%$, Alfa Aesar), boron ($99.9999\%$, Alfa Aesar), and iodine ($99.99\%$, Sigma-Aldrich) were loaded into the quartz tube inside a glovebox with oxygen and water concentrations maintained below 0.1 ppm.
The tube was subsequently evacuated to approximately $10^{-4}$ Torr and sealed.
Third, the sealed tube was placed in a two-zone tube furnace, with the source and growth zones maintained at 1173 and 1123 K, respectively, for 10--14 days.
Finally, the tube was opened after the growth process, and the resulting BAs single crystals were collected for subsequent characterization and measurements.
The high purity of the arsenic source is important for suppressing Si, C, and O impurities in the resulting BAs crystals and minimizing the associated reduction in thermal conductivity caused by point defects~\cite{chen2021effects}.

\subsection*{Experimental parameters}
The incident x-rays were monochromatized by a Si(111) double-crystal monochromator and delivered at a repetition rate of 120 Hz.
The optical pump pulses had a duration of 50 fs and an incident fluence of 22 mJ/cm$^2$. 
The spot size (full width at half maximum) of the optical pump was 160~$\mu$m $\times$ 140~$\mu$m.
The scattered x-rays were recorded using an area detector with a pixel size of 75~$\mu$m $\times$ 75~$\mu$m and an active area of 1000 $\times$ 1000 pixels.
Pulse-to-pulse variations in the incident x-ray intensity were monitored using an ePix100 detector~\cite{sikorski2016application} that measured x-ray scattering from a Kapton window, and the resulting signal was used for normalization.
The relative timing jitter between the optical pump and x-ray probe was corrected on a shot-by-shot basis~\cite{harmand2013achieving}.
The overall experimental temporal resolution was approximately 80~fs.

\subsection*{Phonon calculations}
First-principles calculations were performed using density functional theory as implemented in the Vienna Ab initio Simulation Package (VASP)~\cite{kresse1996efficiency}, with the projector augmented-wave method~\cite{blochl1994projector}.
Structural optimization used the Perdew--Burke--Ernzerhof generalized gradient approximation (PBE-GGA)~\cite{perdew1996generalized}, a plane-wave energy cutoff of 520~eV, and a $\Gamma$-centered $5\times5\times5$ $k$-point mesh.
The total-energy and force convergence criteria were $1\times10^{-7}$~eV and $1\times10^{-3}$~eV\,\AA$^{-1}$, respectively.

Harmonic interatomic force constants (IFCs) were obtained using the finite-displacement method~\cite{parlinski1997first} as implemented in Phonopy~\cite{togo2015first,togo2023first}.
Atomic displacements of $0.01$~\AA{} were introduced in a $5\times5\times5$ supercell of the primitive cell (250 atoms), with a $\Gamma$-centered $1\times1\times1$ $k$-point mesh used for the force calculations.
The resulting IFCs were used to construct and diagonalize the dynamical matrices, yielding the phonon frequencies and eigenvectors, which were subsequently used to calculate the phonon dispersions and one-phonon structure factors.

\section*{References}
\bibliography{references.bib}

\begin{acknowledgments}
We thank Mariano Trigo for the valuable assistance with the experiment and helpful discussions.
This work is based on research supported by the U.S. National Science Foundation (NSF) Future of Semiconductors (FuSe) Program under the award number DMR-2425439. The contributions of H.L., T.S., T.M.L., Y.S., M.H., and D.Z. are supported by the U.S. Department of Energy, Office of Science, Office of Basic Energy Sciences under Contract No. DE-AC02-76SF00515. Use of the Linac Coherent Light Source (LCLS), SLAC National Accelerator Laboratory, is supported by the U.S. Department of Energy, Office of Science, Office of Basic Energy Sciences under Contract No. DE-AC02-76SF00515. Use was also made of computational facilities purchased with funds from NSF (award number CNS-1725797) and administered by the Center for Scientific Computing (CSC) at the University of California, Santa Barbara (UCSB). The CSC is supported by the California NanoSystems Institute and the Materials Research Science and Engineering Center (MRSEC; NSF DMR-2308708) at UCSB.
\end{acknowledgments}

\end{document}



\title{Supplementary Information: Slow Thermalization and Long-Lived Coherence across Acoustic Phonon Branches in BAs} 

\author{Zeyu Xiang}
\thanks{These authors contributed equally.}
\affiliation{Department of Mechanical Engineering, University of California, Santa Barbara, CA 93106, USA}

\author{Ying Peng}
\thanks{These authors contributed equally.}
\affiliation{Department of Physics and Texas Center for Superconductivity at the University of Houston (TcSUH), Houston, TX 77204, USA}

\author{Fanghao Zhang}
\affiliation{Department of Mechanical Engineering, University of California, Santa Barbara, CA 93106, USA}

\author{Haoyuan Li}
\affiliation{SLAC National Accelerator Laboratory, Menlo Park, CA 94025, USA}

\author{Takahiro Sato}
\affiliation{SLAC National Accelerator Laboratory, Menlo Park, CA 94025, USA}

\author{Thomas Linker}
\affiliation{SLAC National Accelerator Laboratory, Menlo Park, CA 94025, USA}
\affiliation{Stanford PULSE Institute, SLAC National Accelerator Laboratory, Menlo Park, CA 94025, USA}

\author{Yanwen Sun}
\affiliation{SLAC National Accelerator Laboratory, Menlo Park, CA 94025, USA}

\author{Meredith Henstridge}
\affiliation{SLAC National Accelerator Laboratory, Menlo Park, CA 94025, USA}

\author{Vincent Esposito}
\affiliation{SLAC National Accelerator Laboratory, Menlo Park, CA 94025, USA}

\author{James F. Cotter}
\affiliation{SLAC National Accelerator Laboratory, Menlo Park, CA 94025, USA}

\author{Diling Zhu}
\affiliation{SLAC National Accelerator Laboratory, Menlo Park, CA 94025, USA}

\author{Zhifeng Ren}
\email{zren@uh.edu} \affiliation{Department of Physics and Texas Center for Superconductivity at the University of Houston (TcSUH), Houston, TX 77204, USA}

\author{Bolin Liao}
\email{bliao@ucsb.edu} \affiliation{Department of Mechanical Engineering, University of California, Santa Barbara, CA 93106, USA}

\maketitle



\section*{Suppmenetary Note 1: Thermal conductivity mapping}

To verify the high quality and thermal conductivity of the synthesized BAs crystals, we performed spatially resolved measurements using frequency-domain thermoreflectance (FDTR)~\cite{schmidt2009frequency}.
Prior to the measurements, the BAs surface was coated with a $5~\mathrm{nm}$ Ti adhesion layer followed by an $80~\mathrm{nm}$ Au transducer layer.
A continuous-wave $445~\mathrm{nm}$ pump laser was amplitude modulated to periodically heat the sample, while a continuous-wave $488~\mathrm{nm}$ probe laser monitored the resulting temperature response through changes in surface reflectance.
The pump and probe spot diameters were approximately $34$ and $3~\mu\mathrm{m}$, respectively.
The phase of the thermoreflectance signal was measured as a function of pump modulation frequency using a lock-in amplifier and fitted to a frequency-domain heat diffusion model to extract the thermal conductivity.

\begin{figure}[!hb]
\includegraphics[width=0.9\textwidth]{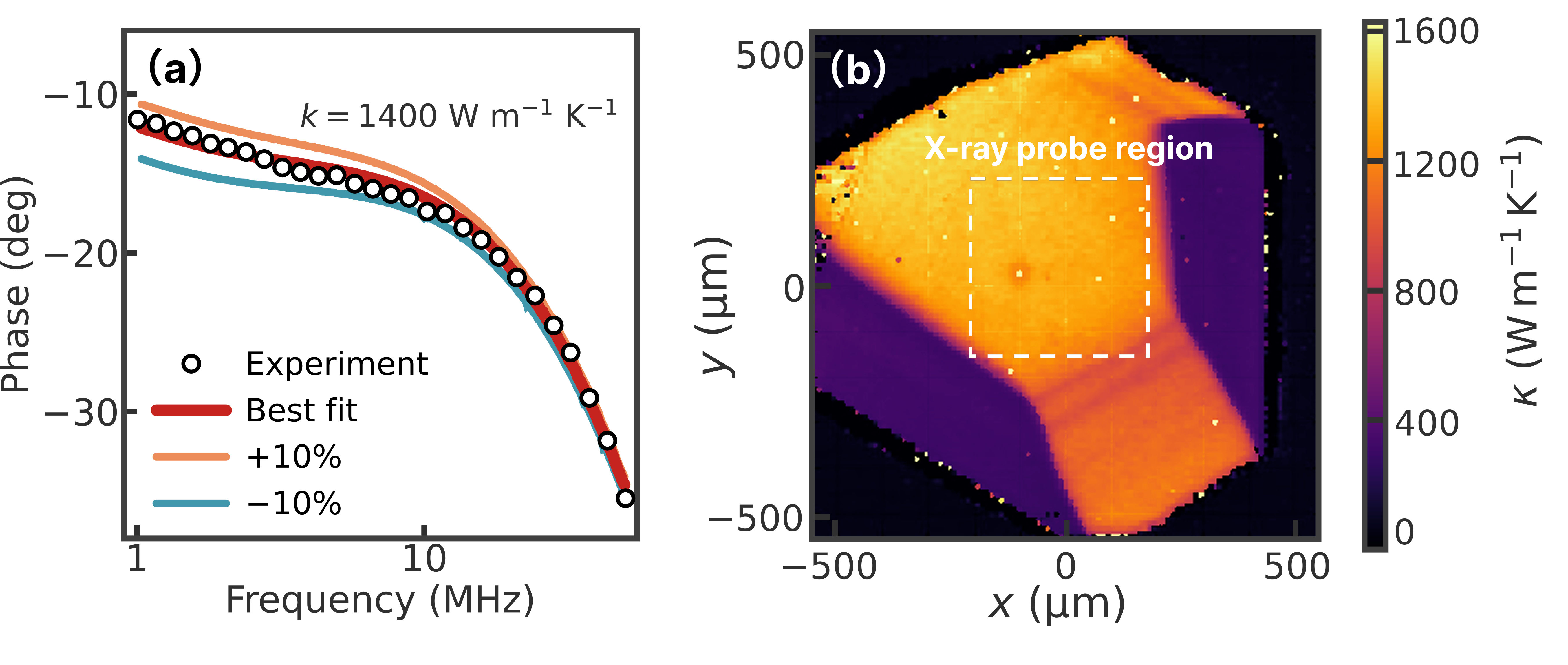}
\caption{(a) Representative FDTR measurement (open circles) and best fit (red line), yielding a thermal conductivity of 1400~$\mathrm{W\,m^{-1}\,K^{-1}}$ at 293 K. The orange and blue curves show the calculated signals for thermal conductivity varied by $+10\%$ and $-10\%$, respectively. (b) Spatial map of the thermal conductivity across the entire BAs sample, showing extensive regions with values ranging from 1400 to 1600~$\mathrm{W\,m^{-1}\,K^{-1}}$.
The dashed box marks the x-ray probe region.
}
\label{fig:figs1}
\end{figure}

Fig.~\ref{fig:figs1}(a) shows a representative FDTR measurement and the corresponding fit, yielding a thermal conductivity of 1400~$\mathrm{W\,m^{-1}\,K^{-1}}$.
We then mapped the thermal conductivity across the entire sample, as shown in Fig.~\ref{fig:figs1}(b), revealing extensive regions with thermal conductivity ranging from 1400 to 1600~$\mathrm{W\,m^{-1}\,K^{-1}}$, confirming the high quality of the sample.
The x-ray probe region used for the time-resolved x-ray diffuse scattering measurements is indicated in Fig.~\ref{fig:figs1}(b).

\clearpage

\section*{Supplementary Note 2: Structure factors across phonon branches}
The one-phonon contribution to the thermal diffuse scattering intensity $I(\mathbf{Q})$ at momentum transfer Q is given by~\cite{xu2005determination,warren1990x}
\begin{equation}
    I(\mathbf{Q}) \propto \sum_j
\frac{n_j(\mathbf{q})+\frac{1}{2}}
{\Omega_j(\mathbf{q})}
\left|F_j(\mathbf{Q})\right|^2,
\label{eq:s1}
\end{equation}
where $\omega_j(\mathbf{q})$ is the angular frequency of phonon branch $j$, and
$n_j(\mathbf{q})
=
\frac{1}
{\exp\!\left[\hbar\Omega_j(\mathbf{q})/(k_{\mathrm B}T_j)\right]-1}$
is the corresponding Bose--Einstein occupation at temperature $T_j$.
$F_j(\mathbf{Q})$ denotes the inelastic structure factor by summing over all atoms in the unit cell~\cite{trigo2010imaging}
\begin{equation}
    F_j(\mathbf{Q})=\sum_s\frac{f_s}{\sqrt{m_s}}e^{-W_s}(\mathbf{Q}\cdot\mathbf{e} _{s,j,\mathbf{q}})e^{-i\mathbf{K}_Q\cdot \mathbf{r}_s},
\label{eq:s2}
\end{equation}
where $f_s$, $m_s$, $\mathbf{r}_s$, and $W_s$ are the atomic scattering factor, mass, equilibrium position, and Debye--Waller factor of atom $s$, respectively.
$\mathbf{K}_Q$ is the reciprocal-lattice vector closest to $\mathbf{Q}$.
The phonon polarization vector $\mathbf{e}_{s,j,\mathbf{q}}$ describes the displacement of atom $s$ in branch $j$ at wave vector $\mathbf{q}$. 
To determine these polarization vectors, the harmonic interatomic force constants (IFCs) are first calculated from first principles using the finite-displacement method~\cite{togo2015first}, after which the dynamical matrices are constructed and diagonalized with Phonopy at the corresponding $\mathbf{q}$ points~\cite{togo2015first,togo2023first}.

For structure factors of BAs, Eq.~\eqref{eq:s2} becomes
\begin{equation}
    F_j(\mathbf{Q})=
    \frac{f_{\mathrm{B}}}{\sqrt{m_{\mathrm{B}}}}
    e^{-W_{\mathrm{B}}}
    (\mathbf{Q}\cdot\mathbf{e}_{\mathrm{B},j,\mathbf{q}})
    e^{-i\mathbf{K}_Q\cdot \mathbf{r}_{\mathrm{B}}}
    +
    \frac{f_{\mathrm{As}}}{\sqrt{m_{\mathrm{As}}}}
    e^{-W_{\mathrm{As}}}
    (\mathbf{Q}\cdot\mathbf{e}_{\mathrm{As},j,\mathbf{q}})
    e^{-i\mathbf{K}_Q\cdot \mathbf{r}_{\mathrm{As}}},
\label{eq:s3}
\end{equation}
where the subscript B and As are added to denote terms associated with boron and arsenic atoms, respectively.
In the long-wavelength limit, acoustic modes correspond to rigid translations of the lattice, for which the B and As atoms move in phase with equal displacements,
\begin{equation}
    \mathbf{u}_{\mathrm{B}}^{\mathrm{ac}}
    =
    \mathbf{u}_{\mathrm{As}}^{\mathrm{ac}}.
\label{eq:s4}
\end{equation}
Since the phonon eigenvector is related to the atomic displacement by
$\mathbf{e}_s \propto \sqrt{m_s}\mathbf{u}_s$~\cite{born1996dynamical}, the normalized acoustic-mode eigenvectors in the long-wavelength limit can be written as
\begin{equation}
    \mathbf{e}_{\mathrm{B}}^{\mathrm{ac}}
    =
    \sqrt{\frac{m_{\mathrm{B}}}{M}}\,\boldsymbol{\xi},
    \qquad
    \mathbf{e}_{\mathrm{As}}^{\mathrm{ac}}
    =
    \sqrt{\frac{m_{\mathrm{As}}}{M}}\,\boldsymbol{\xi},
\label{eq:s5}
\end{equation}
where $M=m_{\mathrm{B}}+m_{\mathrm{As}}$ and $\boldsymbol{\xi}$ denotes the phonon polarization direction.
Substituting these eigenvectors into Eq.~\eqref{eq:s3}, the acoustic-mode structure factor becomes
\begin{equation}
    F_{\mathrm{ac}}(\mathbf{Q})
    =
    \frac{\mathbf{Q}\cdot\boldsymbol{\xi}}{\sqrt{M}}
    \left[
        f_{\mathrm{B}}e^{-W_{\mathrm{B}}}
        e^{-i\mathbf{K}_{Q}\cdot\mathbf{r}_{\mathrm{B}}}
        +
        f_{\mathrm{As}}e^{-W_{\mathrm{As}}}
        e^{-i\mathbf{K}_{Q}\cdot\mathbf{r}_{\mathrm{As}}}
    \right].
\label{eq:s6}
\end{equation}
For comparison, optical modes correspond to out-of-phase motion of the B and As sublattices, while the center of mass remains stationary
\begin{equation}
m_{\mathrm B}\mathbf{u}_{\mathrm B}^{\mathrm{op}}
+
m_{\mathrm{As}}\mathbf{u}_{\mathrm{As}}^{\mathrm{op}}
=0.
\label{eq:s7}
\end{equation}
Therefore, the normalized optical-mode eigenvectors can be written as
\begin{equation}
    \mathbf{e}_{\mathrm{B}}^{\mathrm{op}}
    =
    \sqrt{\frac{m_{\mathrm{As}}}{M}}\,\boldsymbol{\xi},
    \qquad
    \mathbf{e}_{\mathrm{As}}^{\mathrm{op}}
    =
    -\sqrt{\frac{m_{\mathrm{B}}}{M}}\,\boldsymbol{\xi}.
\label{eq:s8}
\end{equation}
Substituting these eigenvectors into Eq.~\eqref{eq:s3}, the optical-mode structure factor becomes
\begin{equation}
    F_{\mathrm{op}}(\mathbf{Q})
    =
    \frac{\mathbf{Q}\cdot\boldsymbol{\xi}}{\sqrt{M}}
    \left[
        f_{\mathrm{B}}
        \sqrt{\frac{m_{\mathrm{As}}}{m_{\mathrm{B}}}}
        e^{-W_{\mathrm{B}}}
        e^{-i\mathbf{K}_{Q}\cdot\mathbf{r}_{\mathrm{B}}}
        -
        f_{\mathrm{As}}
        \sqrt{\frac{m_{\mathrm{B}}}{m_{\mathrm{As}}}}
        e^{-W_{\mathrm{As}}}
        e^{-i\mathbf{K}_{Q}\cdot\mathbf{r}_{\mathrm{As}}}
    \right].
\label{eq:s9}
\end{equation}
Compared with the acoustic modes, the opposite-phase sublattice motion introduces an opposite sign between the B and As contributions to the optical-mode structure factor, enabling destructive interference depending on their crystallographic phase relation.

We choose the (220) Bragg reflection because it provides a large acoustic-to-optical structure-factor contrast, thereby enhancing the sensitivity to acoustic phonons.
For zinc-blende BAs, the two atoms in the primitive cell are located at
\begin{equation}
    \mathbf{r}_{\mathrm{B}}
    =
    (0,0,0),
    \qquad
    \mathbf{r}_{\mathrm{As}}
    =
    \frac{a}{4}(1,1,1),
\label{eq:s10}
\end{equation}
where $a$ is the conventional cubic lattice constant.
For momentum transfers near the conventional cubic (220) reflection,
$\mathbf{K}_Q=\mathbf{G}_{220}$, with
$\mathbf{G}_{220}=\mathbf{b}_1+\mathbf{b}_2+2\mathbf{b}_3$
in the primitive reciprocal basis.
The corresponding relative phase factor between the B and As sublattices is
\begin{equation}
    e^{-i\mathbf{G}_{220}\cdot
    (\mathbf{r}_{\mathrm{As}}-\mathbf{r}_{\mathrm{B}})}
    =
    1.
\label{eq:s11}
\end{equation}
Thus, the B and As sublattices acquire the same crystallographic phase factor at the (220) reflection.
The optical-mode structure factor in Eq.~\eqref{eq:s9} therefore reduces to
\begin{equation}
    F_{\mathrm{op}}^{(220)}(\mathbf{Q})
    =
    \frac{\mathbf{Q}\cdot\boldsymbol{\xi}}{\sqrt{M}}
    e^{-i\mathbf{G}_{220}\cdot\mathbf{r}_{\mathrm{B}}}
    \left[
        f_{\mathrm{B}}
        \sqrt{\frac{m_{\mathrm{As}}}{m_{\mathrm{B}}}}
        e^{-W_{\mathrm{B}}}
        -
        f_{\mathrm{As}}
        \sqrt{\frac{m_{\mathrm{B}}}{m_{\mathrm{As}}}}
        e^{-W_{\mathrm{As}}}
    \right].
\label{eq:s12}
\end{equation}
For a simple estimate, we neglect the Debye--Waller attenuation by taking
$e^{-W_{\mathrm{B}}}\approx e^{-W_{\mathrm{As}}}\approx1$ and approximate the x-ray atomic scattering factors by the corresponding atomic numbers,
$f_{\mathrm{B}}\approx Z_{\mathrm{B}}=5$ and
$f_{\mathrm{As}}\approx Z_{\mathrm{As}}=33$~\cite{waasmaier1995new}.
The two mass-weighted contributions in Eq.~\eqref{eq:s12} are then
\begin{equation}
    Z_{\mathrm{B}}
    \sqrt{\frac{m_{\mathrm{As}}}{m_{\mathrm{B}}}}
    \approx 13.2,
    \qquad
    Z_{\mathrm{As}}
    \sqrt{\frac{m_{\mathrm{B}}}{m_{\mathrm{As}}}}
    \approx 12.5.
\end{equation}
Their comparable magnitudes lead to strong destructive interference, substantially suppressing the optical-mode structure factor near the (220) reflection.
Importantly, this suppression is reflection dependent.
For a general $(\mathrm{hkl})$ reflection, the relative crystallographic phase between the B and As sublattices is not necessarily unity, and their contributions to the optical-mode structure factor therefore do not necessarily interfere destructively.
For example, for the $(33\bar{1})$ reflection,
\begin{equation}
    e^{-i\mathbf{G}_{33\bar{1}}\cdot
    (\mathbf{r}_{\mathrm{As}}-\mathbf{r}_{\mathrm{B}})}
    =
    e^{-i5\pi/2}
    =
    -i,
\label{eq:s15}
\end{equation}
such that the two optical-mode contributions are no longer directly out of phase.

\begin{figure}[!htb]
\includegraphics[width=0.9\textwidth]{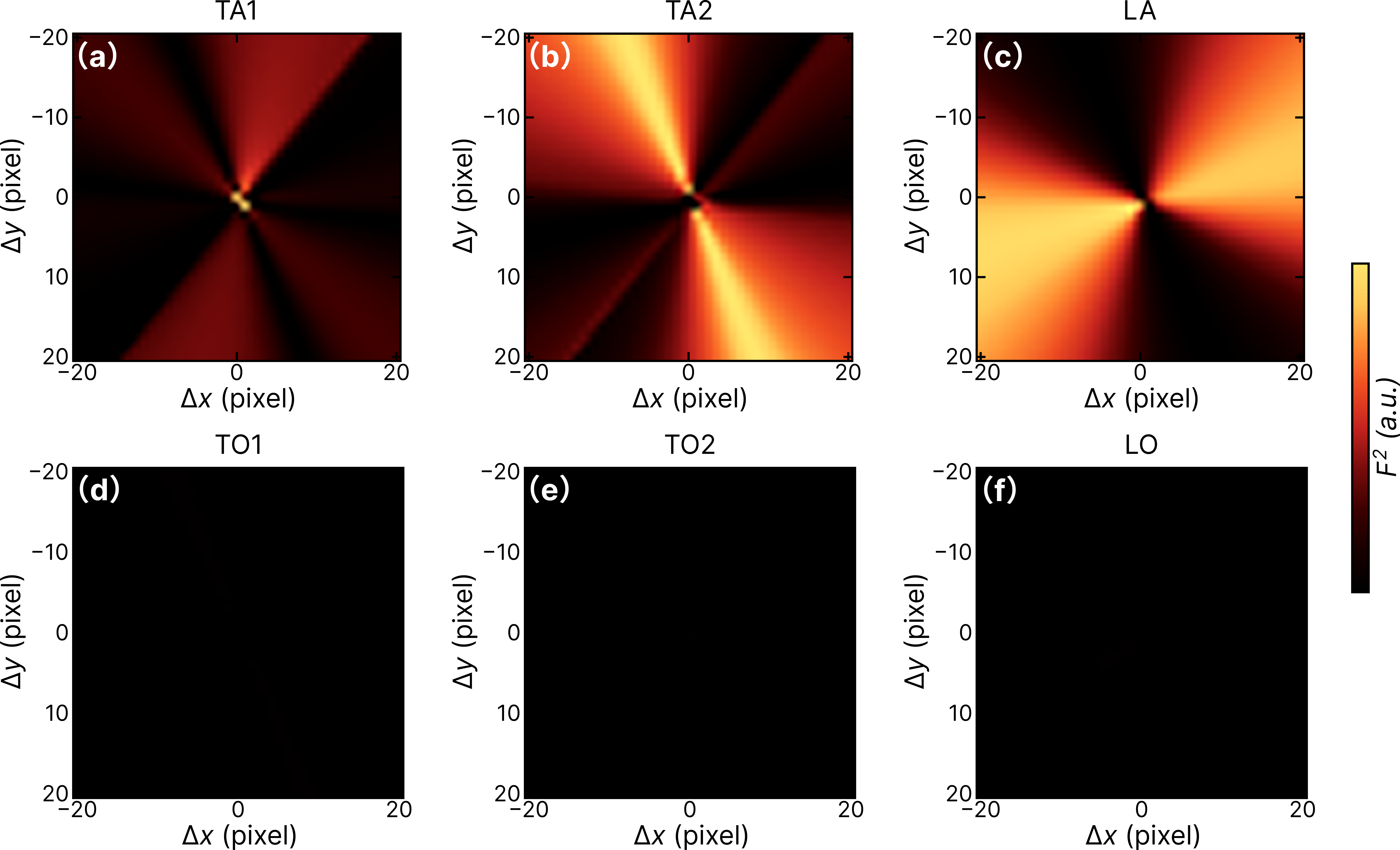}
\caption{One-phonon inelastic structure-factor intensities calculated over the experimentally sampled $40\times40$-pixel reciprocal-space region near the (220) Bragg peak for (a)--(c) the acoustic branches and (d)--(f) the optical branches.}
\label{fig:figs2}
\end{figure}

Based on the phonon wave vectors $\mathbf{q}$ sampled within the region of interest near the (220) Bragg peak, we calculate the one-phonon inelastic structure factors for the different acoustic branches, as shown in Fig.~\ref{fig:figs2}.
The acoustic branches exhibit strongly anisotropic and branch-dependent intensity distributions.
In contrast, the calculated structure factors of the three optical branches are substantially weaker over the same reciprocal-space region and are nearly invisible on the common intensity scale.
This strong contrast indicates that the measured diffuse scattering response near the (220) reflection is dominated by acoustic phonons.

\clearpage

\section*{Supplementary Note 3: Rocking Curve and Sample Detuning}

\begin{figure}[!htb]
\includegraphics[width=0.45\textwidth]{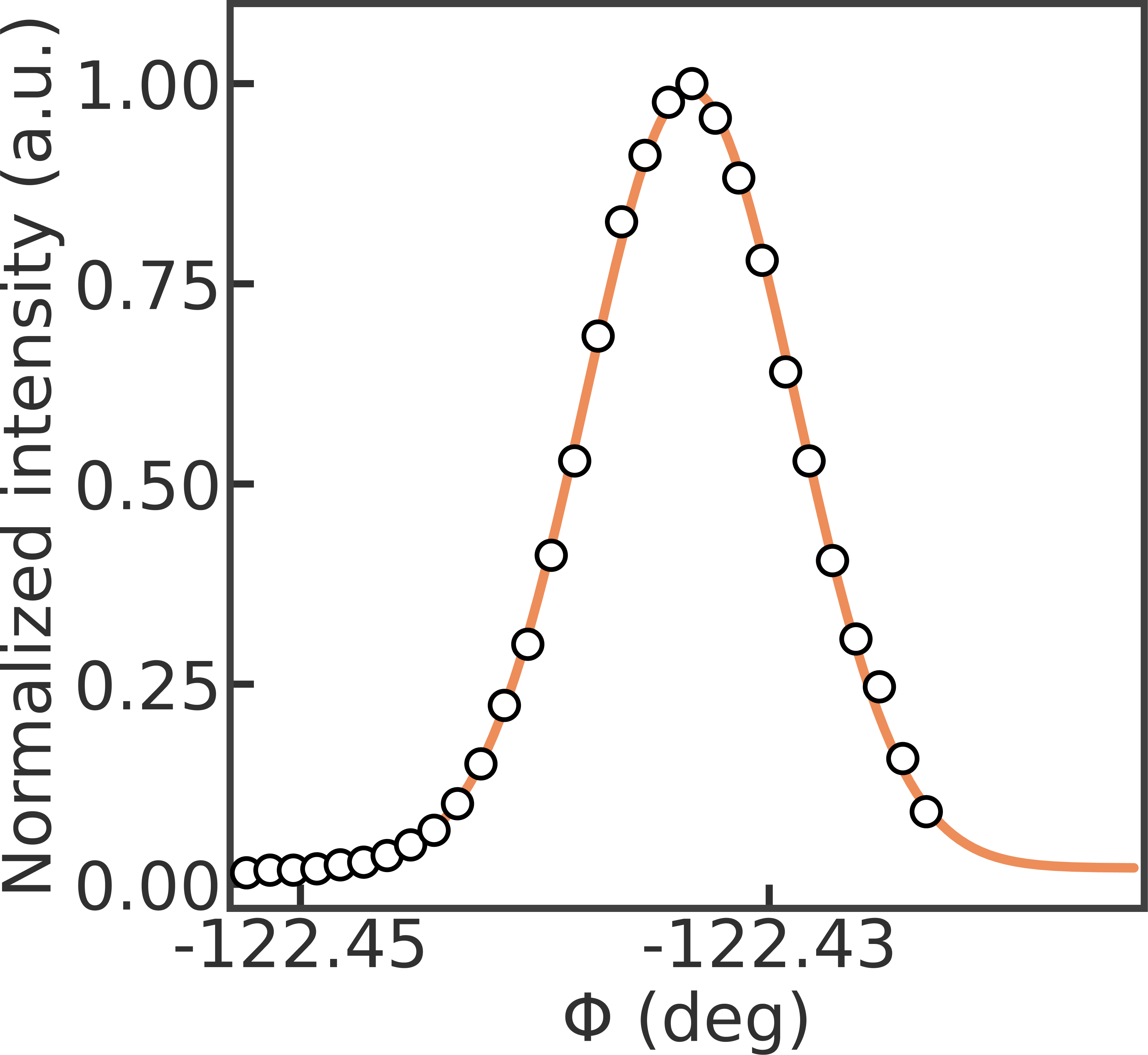}
\caption{Rocking curve measured near the (220) Bragg reflection.
Open circles show the experimental data, and the orange line represents the fit.
}
\label{fig:figs3}
\end{figure}

To determine the exact Bragg condition of the (220) reflection, we measured a rocking curve by varying the sample angle $\Phi$ about the surface normal (the geometry is shown in Fig.~1(a) in the main text).
As shown in Fig.~\ref{fig:figs3}, the measured intensity exhibits a well-defined maximum, defining the exact Bragg condition.
For the time-resolved measurements, the sample was subsequently detuned by $\Delta\Phi=0.2^\circ$ from this position.
The $0.2^\circ$ detuning is substantially larger than the approximately $0.011^\circ$ full width at half maximum of the (220) rocking curve, thereby strongly suppressing the Bragg contribution while retaining sensitivity to the surrounding diffuse scattering used to probe the phonon dynamics.

\clearpage

\section*{Supplementary Note 4: Momentum Dependence of the Buildup Time}
\begin{figure}[!htb]
\includegraphics[width=0.9\textwidth]{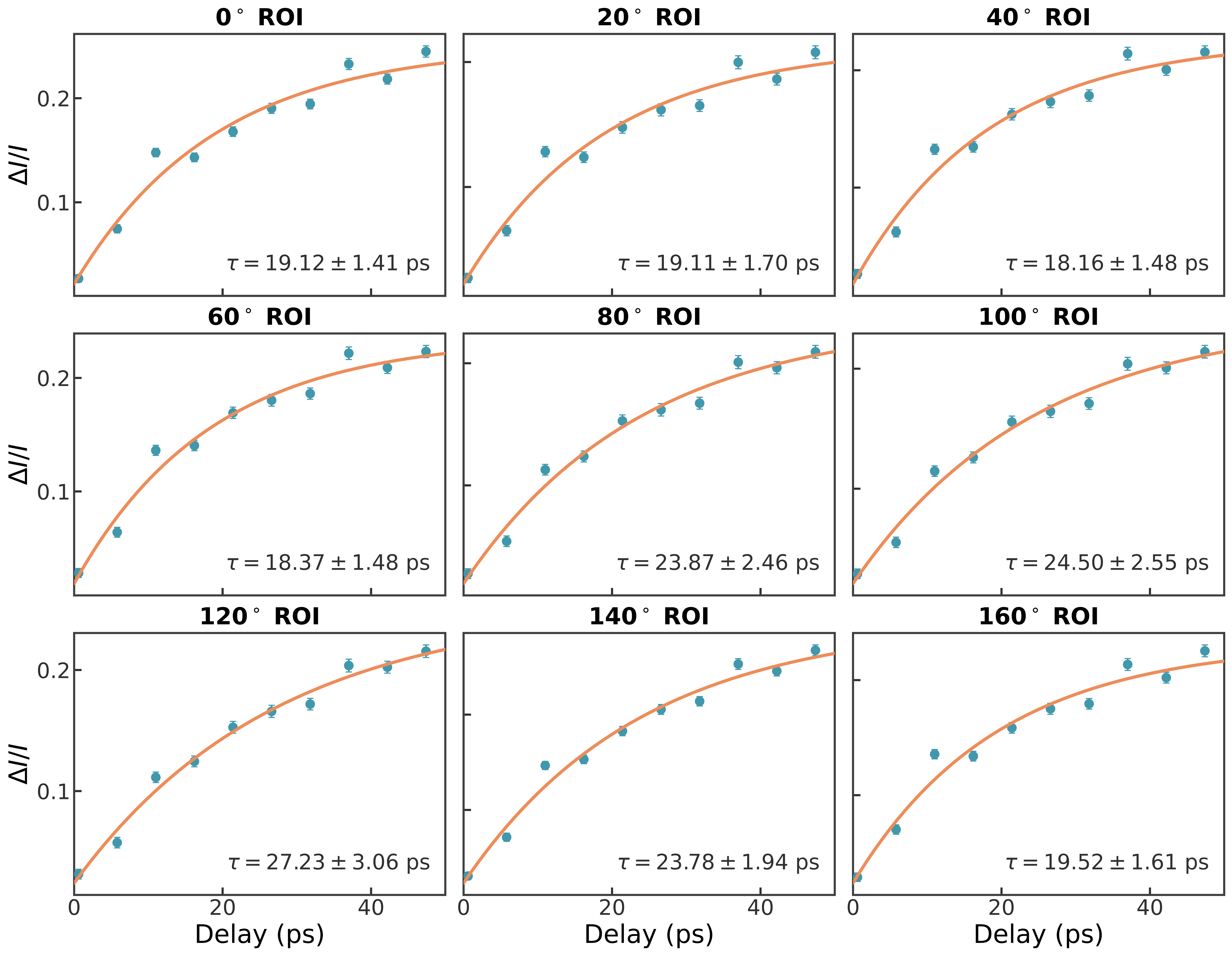}
\caption{Momentum-dependent buildup dynamics near the (220) Bragg reflection.
Blue circles with error bars show the measured diffuse scattering intensity, and the orange curves represent single-exponential fits.
The corresponding buildup time $\tau(\theta)$ is indicated in each panel.
}
\label{fig:figs4}
\end{figure}

To quantify the momentum dependence of the diffuse scattering buildup, we analyzed the temporal response along each momentum direction around the (220) Bragg reflection (as shown in Fig.~2(d) in the main text), with the data binned using a temporal interval of $\Delta t=5$ ps.
For each direction, the measured intensity was fitted with a single exponential rise,
\begin{equation}
\frac{\Delta I}{I}(t)
=
C+A\left(1-e^{-t/\tau}\right),
\end{equation}
where $C$ is an offset, $A$ is the amplitude of the intensity increase, and $\tau$ is the effective buildup time.
Fig.~\ref{fig:figs4} shows the measured responses and corresponding fits for all momentum directions.
The single exponential model captures the overall buildup of the diffuse scattering, while the extracted $\tau$ varies systematically with momentum direction.
These values are used to construct the angular dependence of the buildup time shown in Fig.~2(e) of the main text.

\clearpage

\section*{Supplementary Note 5: Acoustic Phonon Dispersion and TA Branch Degeneracy}
\begin{figure}[!htb]
\includegraphics[width=0.9\textwidth]{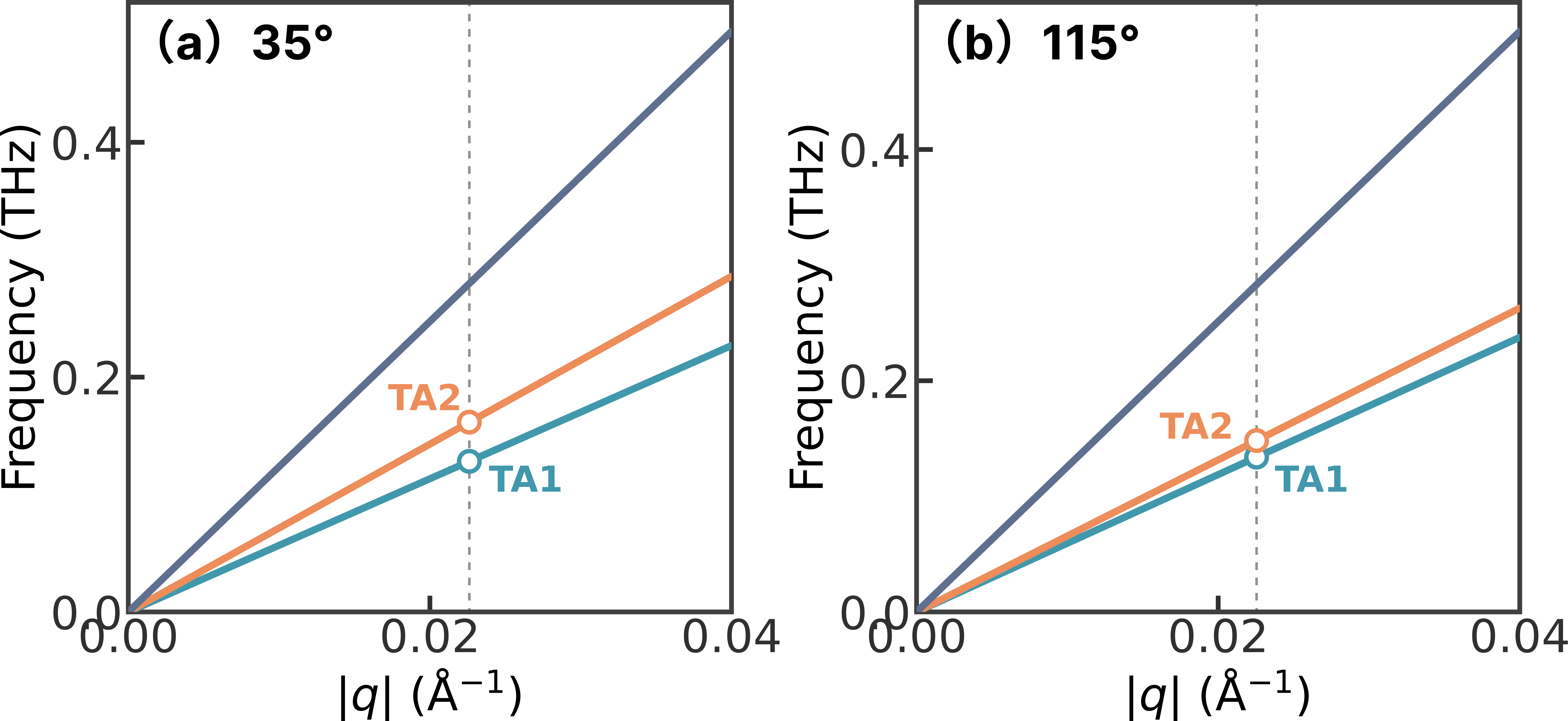}
\caption{
(a) Phonon dispersion along the $\theta=35^\circ$ direction, corresponding to the normalized momentum direction $\hat{\mathbf{q}}\approx(-0.76,0.04,-0.65)$. At the experimentally sampled wavevector $\mathbf{q}\approx(-0.017,0.001,-0.015)~\mathrm{\AA}^{-1}$, the two transverse acoustic branches have frequencies $f_{\mathrm{TA1}}\approx0.13$ THz and $f_{\mathrm{TA2}}\approx0.16$ THz. (b) Phonon dispersion along the $\theta=115^\circ$ direction, corresponding to $\hat{\mathbf{q}}\approx(0.40,0.57,-0.71)$. At $\mathbf{q}\approx(0.009,0.013,-0.016)~\mathrm{\AA}^{-1}$, the corresponding TA frequencies are $f_{\mathrm{TA1}}\approx0.13$ THz and $f_{\mathrm{TA2}}\approx0.15$ THz.
}
\label{fig:figs5}
\end{figure}

Fig.~\ref{fig:figs5} shows the first-principles phonon dispersions calculated within the same framework used above to obtain the phonon eigenvectors, along the $\theta=35^\circ$ and $115^\circ$ ROI directions.
The corresponding normalized momentum directions are $\hat{\mathbf{q}}\approx(-0.76,0.04,-0.65)$ and $(0.40,0.57,-0.71)$, with representative reduced wavevectors $\mathbf{q}\approx(-0.017,0.001,-0.015)~\mathrm{\AA}^{-1}$ and $(0.009,0.013,-0.016)~\mathrm{\AA}^{-1}$, respectively.
At these wavevectors, the calculated TA1 and TA2 frequencies are approximately 0.13 and 0.16 THz at $\theta=35^\circ$, and 0.13 and 0.15 THz at $\theta=115^\circ$, while the LA frequency is approximately 0.28 THz in both directions.
The resulting TA branch splittings of 0.03 and 0.02 THz are much smaller than the TA--LA separation, supporting the treatment of the two TA branches as a combined TA contribution.

\clearpage

\section*{Supplementary Note 6: Theoretical Acoustic Phonon Thermalization}
\begin{figure}[!htb]
\includegraphics[width=0.45\textwidth]{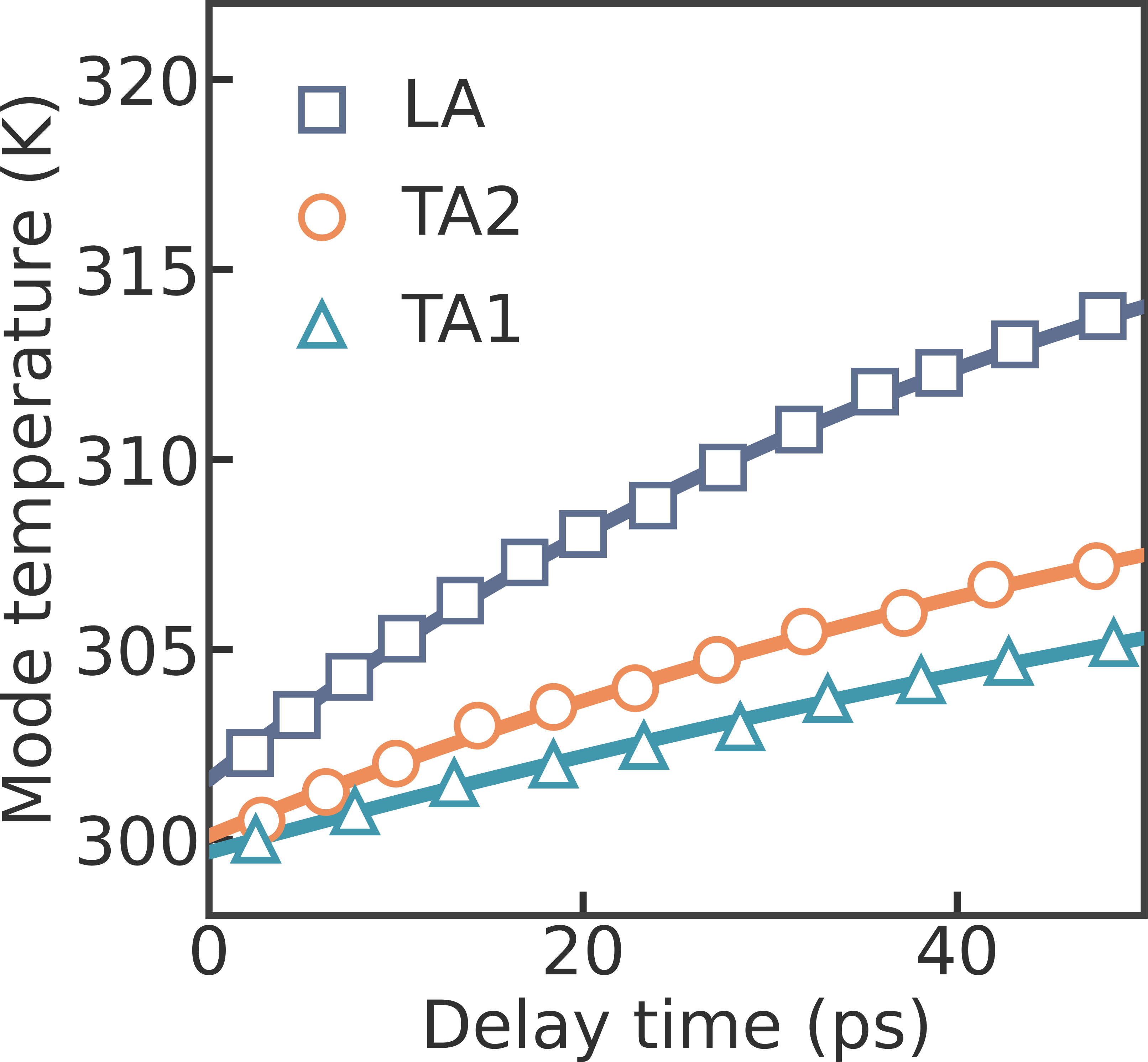}
\caption{
Branch averaged mode temperatures for the LA, TA1, and TA2 acoustic phonons extracted from the calculations reported in Ref.~\cite{liu2026higher}.
Symbols represent the digitized data, and solid lines show exponential fits performed here, yielding thermalization times of $51.7\pm1.1$ ps for LA, $75.2\pm4.5$ ps for TA2, and $121.8\pm10.5$ ps for TA1.
}
\label{fig:figs6}
\end{figure}

To compare the experimentally observed branch dependent thermalization with theoretical predictions, we extracted the branch averaged acoustic phonon temperatures reported in Ref.~\cite{liu2026higher}.
As shown in Fig.~\ref{fig:figs6}, the LA branch heats more rapidly than the two TA branches following photoexcitation.
Fitting the extracted temperature evolution with the same exponential form used for the experimental data yields thermalization times of $51.7\pm1.1$ ps for LA, $75.2\pm4.5$ ps for TA2, and $121.8\pm10.5$ ps for TA1.
The corresponding TA2/LA and TA1/LA timescale ratios are approximately 1.45 and 2.36, respectively, bracketing the experimentally measured combined TA/LA ratio of approximately 2.10.
The absolute theoretical and experimental timescales are not expected to coincide quantitatively because the calculations describe branch averaged phonon populations under different excitation conditions, whereas the experiment probes structure factor weighted populations within a restricted small-$q$ region near the Brillouin zone center.

\clearpage

\section*{Supplementary Note 7: Early-time scattering dynamics}
\begin{figure}[!htb]
\includegraphics[width=0.45\textwidth]{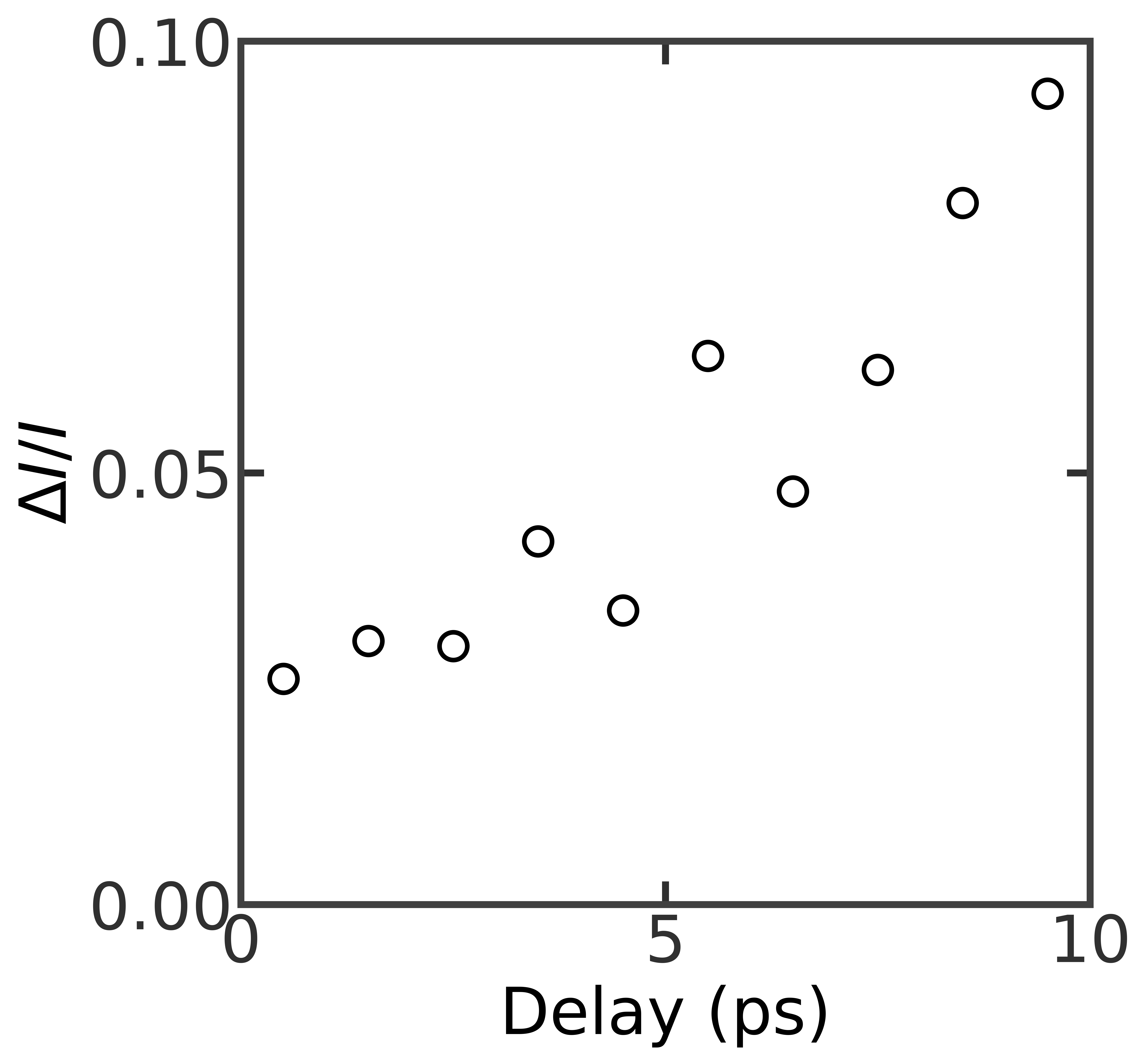}
\caption{Measured $\Delta I/I$ response during the first 10~ps following photoexcitation.
The signal is dominated by the initial buildup, with no well-defined periodic oscillation resolved within this time window.}
\label{fig:figs7}
\end{figure}

Fig.~\ref{fig:figs7} shows the measured $\Delta I/I$ response during the first 10~ps following photoexcitation.
The signal is dominated by the initial buildup, with no well-defined periodic oscillation resolved within this time window.
The pronounced coherent oscillations become apparent at later delays, as shown in Fig.~3 of the main text.

\clearpage

\section*{Supplementary Note 8: Coherent Oscillation Frequency Robustness against Temporal Binning}
\begin{figure}[!htb]
\includegraphics[width=0.9\textwidth]{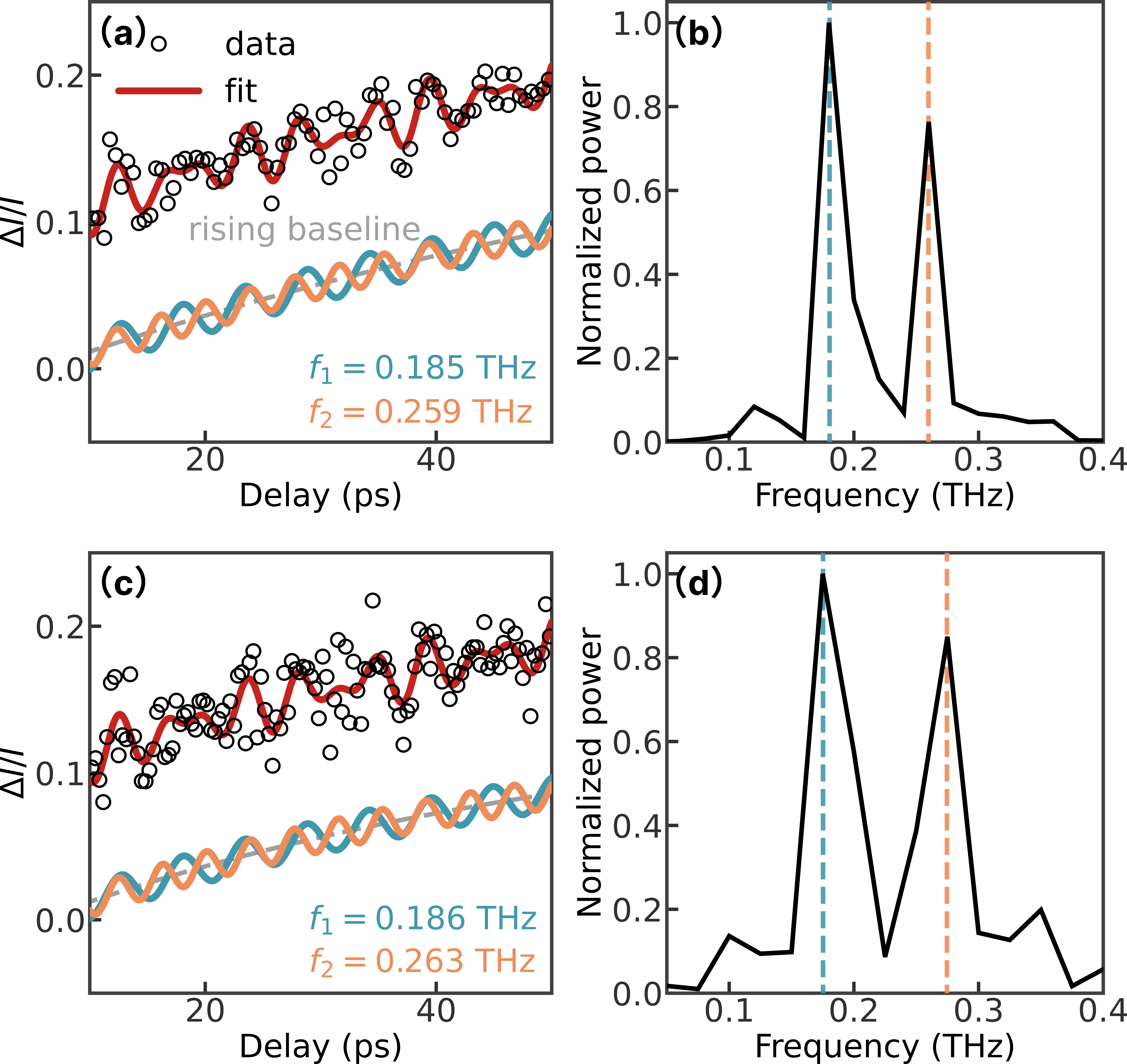}
\caption{
(a) Representative $\Delta I/I$ transient obtained with a temporal bin width of $\Delta t=0.5$ ps, together with a fit comprising a slowly rising background and two oscillatory components. The blue and orange curves show the extracted components at $f_1=0.185$ THz and $f_2=0.259$ THz, respectively.
(b) Corresponding normalized Fourier power spectrum, showing peaks near the two fitted frequencies.
(c) Same analysis using a temporal bin width of $\Delta t=0.25$ ps, yielding $f_1=0.186$ THz and $f_2=0.263$ THz.
(d) Corresponding normalized Fourier power spectrum.
The persistence of both frequency components across the two bin widths demonstrates the robustness of the extracted oscillation frequencies.
}
\label{fig:figs8}
\end{figure}

To assess the sensitivity of the extracted oscillation frequencies to temporal binning, we repeated the analysis using bin widths of $\Delta t=0.5$ and $0.25$ ps.
As shown in Fig.~\ref{fig:figs8}, both frequency components remain clearly resolved for the two bin widths.
The fits yield $f_1=0.185$ THz and $f_2=0.259$ THz for $\Delta t=0.5$ ps, and $f_1=0.186$ THz and $f_2=0.263$ THz for $\Delta t=0.25$ ps.
The corresponding Fourier spectra exhibit peaks at consistent frequencies, confirming that the extracted oscillation frequencies are robust against the choice of temporal bin width.

\clearpage

\section*{Supplementary Note 8: Wavelet Analysis of the Coherent Oscillations}
To characterize the coherent oscillations in both the time and frequency domains, we perform a continuous Morlet-wavelet transform of the normalized diffuse scattering intensity by~\cite{torrence1998practical}
\begin{equation}
W(s,\tau)
=
\frac{1}{\sqrt{s}}
\int_{-\infty}^{\infty}
\frac{\Delta I(t)}{I}
\,
\psi^{*}
\left(
\frac{t-\tau}{s}
\right)
dt,
\end{equation}
where $s$ is the wavelet scale, $\tau$ is the time translation, and the asterisk denotes complex conjugation. We use the complex Morlet mother wavelet,
\begin{equation}
\psi(\eta)
=
\pi^{-1/4}
\exp(i\omega_0\eta)
\exp\left(-\frac{\eta^2}{2}\right),
\end{equation}
where $\eta=(t-\tau)/s$ and $\omega_0$ is the dimensionless central frequency. The corresponding wavelet power is given by
\begin{equation}
P(s,\tau)
=
\left|W(s,\tau)\right|^2.
\end{equation}
For the analysis in this paper, the transient signals are binned with a temporal interval of $\Delta t=0.5$ ps and analyzed over the delay range $\tau=10$--$50$ ps. 
We set the dimensionless central frequency of the Morlet wavelet to $\omega_0=6$ and evaluate wavelet scales corresponding to frequencies between 0.1 and 0.4 THz.
The wavelet scale is converted to the equivalent Fourier frequency according to
\begin{equation}
f(s)
=
\frac{\omega_0+\sqrt{2+\omega_0^2}}
{4\pi s}.
\end{equation}
For $\omega_0=6$, the analyzed frequency range corresponds to wavelet scales of approximately $s=2.42$--$9.68$ ps.

\clearpage

\section*{Supplementary References}

\bibliography{references.bib}